\documentclass[aps,prb,twocolumn,byrevtex]{revtex4-2}

 \usepackage{hyperref}
 \usepackage{epsfig}
 \usepackage{color}
 \usepackage{amsmath}
 \usepackage{amsthm}
 \usepackage{amsfonts}
 \usepackage{amssymb}
 \usepackage{orcidlink}

 \usepackage[english]{babel}
 \usepackage[utf8]{inputenc}

\def\be{\begin{equation}}
\def\eea{\end{eqnarray}}
\def\ee{\end{equation}}
\def\bea{\begin{eqnarray}}
\def\ea{\end{array}}
\def\ba{\begin{array}}

\newcommand{\exval}[1]{\mbox{$\left\langle \, {#1}\, \right\rangle$}}

\newcommand{\bel}[1]{\begin{equation}\label{#1}}

\def\zzz{{\mathchoice {\hbox{$\sf\textstyle Z\kern-0.4em Z$}}
{\hbox{$\sf\scriptstyle Z\kern-0.3em Z$}}
{\hbox{$\sf\scriptscriptstyle Z\kern-0.2em Z$}}
{\hbox{$\sf\textstyle Z\kern-0.4em Z$}}}}

\begin{document}

\title{A fermionic path integral for \\ exact enumeration of
  polygons on the 
  simple cubic lattice}

\author{G. M. Viswanathan
\orcidlink{0000-0002-2301-5593 }}

\affiliation{Department of Physics
and \mbox{National Institute of Science and Technology of Complex Systems,}
Federal University of Rio Grande do Norte,
59078-970
Natal--RN, Brazil}


\begin{abstract}

Enumerating polygons on regular lattices is a classic problem in
rigorous statistical mechanics.
The goal of enumerating polygons on the square lattice via fermionic
path integration was achieved using a free-fermion
quadratic action
in the late 1970s.
Given that  polygon edges  only link 2  vertices,
it is {considered plausible, if not natural,} that  
an action of degree 2 in the Grassmann variables
might suffice to enumerate lattice polygons in any dimension.
Nevertheless, on nonplanar lattices the problem has remained open for
more than four decades. Here we derive the Grassmann action for exact
enumeration of polygons on the simple cubic lattice. Moreover, we
prove that this action is not quadratic but quartic --- corresponding
to a model of interacting fermions.

\end{abstract}

\maketitle

\section{Introduction}

An important problem in statistical mechanics and enumerative
combinatorics concerns how to count certain kinds of
objects, such as polygons, polinominoes and polycubes of fixed size,
on regular lattices~\cite{ppp}.  Here we address the problem of exact
enumeration of multipolygons, defined as  connected or disconnected
undirected 
graphs
whose edges can link only nearest-neighbor
lattice 
sites
and all of whose nodes have even degree.
A  free-fermion quadratic Grassmann action for exact enumeration of
2D multipolygons
on
the square lattice has been known
since the late 1970s~\cite{samuel1978}. These  results were
subsequently 
published by Samuel~\cite{samuel1,samuel2,samuel3}
 in 1980 in a seminal trilogy of
 papers
 presenting the fermionic path
integral formulation of classical
lattice spin
models.
Many advances
followed~\cite{sI-2022,sI-01,sI-02,sI-03,sI-2020,sIII-01,sIII-03,sIII-02,grassmanization,sI-04,plechko,y1991,polyakov-1987},
yet for the next 42 years
the analogous problem of finding the action for enumerating 3D
multipolygons had remained unsolved --- until now.
Here we derive the Grassmann action for exact enumeration of
multipolygons on the simple cubic lattice.

Moreover, considering that an edge connects precisely two vertices of
a polygon and assuming that both vertices contribute one Grassmann
variable each to the edge terms of the action, it follows that a
quadratic Grassmann action should suffice for polygon enumeration not
only in two dimensions but in any lattice dimension.  
{The underlying intuition is rooted in familiar commonplace 
  observations, for example if one wishes to string together a bead
  necklace, it suffices to make 2 holes per bead.}
Indeed, 
it is often  assumed that polygons on the cubic lattice should
be enumerable with a quadratic action.
%
This conjecture, though plausible,
turns out to be wrong.  Overturning the conventional wisdom, we prove
that the action that enumerates polygons on the simple cubic lattice
is not quadratic but quartic --- hence unsolvable via existing methods
that rely on the usual pfaffian methods that apply to quadratic
actions.
We also show that, quite remarkably, the
action for the cubic lattice is not polynomial in the edge
weights.

Our strategy is to exploit how the ferromagnetic 3D Ising model can be
formulated in two different ways.  On the one hand, it can be
formulated in a low temperature variable in terms of the generating
function of closed surfaces on the simple cubic
lattice~\cite{samuel3}.  On the other hand, it is also possible to
formulate the model in a high temperature variable in terms of the
generating function of multipolygons. The latter approach is more
commonly used in statistical
mechanics~\cite{feynman-statmech,mccoy-2014,hansel,guttmann-enting}.
We leverage the known~\cite{guttmann-enting} exact correspondence
between the two formulations and ``work backwards,'' thereby obtaining
the desired Grassmann action.

 { The high and low temperature formulations of the
  Ising model are reviewed in Sec.~\ref{hilo}.  A review of Grassmann
  variables and Berezin integration is given in
  Sec.~\ref{sec-action}. This section also gives a detailed
  explanation of the Grassmann actions for the 2D and 3D Ising model.
  Sec.~\ref{sec-results} presents the results and advances and 
  Sec.~\ref{sec-end} concludes with a brief discussion.  }



\section
{The high and low temperature formulations of the 3D Ising model.}
\label{hilo}

The Hamiltonian of the  isotropic  ferromagnetic Ising model
with $N$ classical spins with 
nearest-neighbor interactions and 
zero external magnetic
field is typically defined as 
$
H_N=
- J \sum_{\exval{ij}}  
\sigma_i \sigma_j
$
where the $J$ is the isotropic coupling, $\sigma_i=\pm 1$ and
$\exval{ij}$ denotes the set of all pairs $(i,j)$ of nearest-neighbor
spins on the chosen lattice.  
The total number of sites is $N=L^d$ for a
system of linear size $L$ on 
a $d$-dimensional grid
lattice.
In what follows we restrict our attention to the cases \mbox{$d=2,3$}
for the square and simple cubic lattices respectively.
%
Although diverse boundary conditions can
be used, we assume periodic boundary conditions without loss of generality.
Indeed,
it
is well known that the thermodynamic limit of the nearest-neighbor
Ising model is independent of boundary conditions.
The canonical partition function is conventionally defined according
to \mbox{$Z_N(\beta)= \sum \exp[-\beta H_N]$,} where
\mbox{$\beta=1/(k_{\mbox{\tiny B}}T)$} is the temperature parameter as usual
and the sum is over all possible spin configurations.


%
Let $K=\beta J$ be the (adimensional) reduced temperature parameter.
For studying the behavior at low temperatures (i.e. high $\beta$), by
convention~\cite{samuel1,gattringer} one uses the low temperature
variable
%
$u=\exp[-K]$
%
(although sometimes the square or fourth power of this quantity is used
instead~\cite{guttmann-enting}).
On the other hand, for studying the high
temperature behavior of the model it is often
preferable~\cite{feynman-statmech,mccoy-2014,hansel,guttmann-enting} 
to use the high temperature variable
%
$t= \tanh[K]$.
%
For non-negative reduced temperatures $0\leq K \leq +\infty$, the low
and high temperature variables are uniquely given by each other,
according to
%
%
\begin{align}
u&= {\sqrt{1-t \over1+t}}  \nonumber\\
t&= {1-u^2 \over 1+u^2}~. \nonumber 
\end{align}

In the high temperature variable $t$, 
 it is
well known~\cite{feynman-statmech} that the partition function can be expressed as
\be
\label{eq-35uhgjtjtj}
Z_N = 2^N (1-t^2)^{-Nd/2} \Lambda_N(t)~, %
\ee
where $\Lambda(t)$ is the generating function for the number of
multipolygons of fixed length on the $d$-dimensional lattice.
Specifically, if we write $\Lambda_N$ as the expansion
%
\[
\Lambda_N(t)= \sum_{n=0}^{Nd} a_n t^{n}~,
\]
%
then $a_n$ is the number of multipolygons with $n$ edges, with the
total number of all edges being $Nd$. Note that $a_0=1$,
corresponding to the single empty graph.
For clarity we repeat here the textbook derivation of 
the Eq. (\ref{eq-35uhgjtjtj}).
First observe that $\sigma_i\sigma_j=\pm 1$, so that 
  $\exp[-\beta H_N]$ is given by
\begin{align}
  \exp\left[K
    \textstyle
    \sum
\sigma_i\sigma_j\right]
  &= \prod \exp[K\sigma_i\sigma_j] \nonumber
\\
  &= \prod \cosh K +  \sigma_i\sigma_j  \sinh K \nonumber
\\
  &= \prod (\cosh K) (1 +  \sigma_i\sigma_j t)  ~.%
\end{align}
Noting that $\cosh K= 1/\sqrt{1-t^2}$ and that the total number of terms
in the product is equal to the number $Nd$  of bonds, we can write the
partition function as
\be
\label{eq245}
Z_N = (1-t^2)^{-Nd/2} \sum_{\sigma_1}\sum_{\sigma_2}\cdots \sum_{\sigma_N} 
\prod_{\small \exval{ij}} (1 +  \sigma_i\sigma_j t)  ~.%
\ee
Each $\sigma_i$ is summed over $\pm 1$, so the only terms that survive
upon expanding the product are those with no odd powers of any of the
$\sigma_i$. Remaining even powers $\sigma_i^{2n}$ contribute a factor
$1^{2n}+ (-1)^{2n}=2$ for each of the $N$ sites.  The product is over
all possible nearest-neighbor ``bonds,'' hence
%
\be
\label{eq-383o4ig}
Z_N = 2^N(1-t^2)^{-Nd/2}
\sum_{G\in \mathcal M} t^{m(G)}
~, %
\ee
where $m(G)$ is the number of edges of the graph $G$ and $\mathcal M$
is the set of all undirected unweighted graphs on the lattice whose
edges only link nearest-neighbor nodes and whose nodes all have even
degree, with the unique empty graph having $m=0$.
Eq. (\ref{eq-35uhgjtjtj}) follows from (\ref{eq-383o4ig}) by observing
that these graphs $G \in \mathcal M$ are precisely what we have been
calling multipolygons.  Feynman~\cite{feynman-statmech} referred to 
multipolygons 
as  ``closed graphs'' due to the fact that there are no dangling
edges or common sides, i.e. no nodes of degree 1 or 3.
In 3D such multipolygons need not be
planar.


Alternatively, we can express the partition function in terms
of the low temperature variable $u$, by counting
the
number of magnetic domain wall configurations. Let $\Xi(u)$ be the
generating function for the number of domain wall configurations of
fixed size~\cite{gattringer}.  In 2D this will again correspond to the
number of multipolygons of fixed length (because the 2D Ising model is
self-dual). In contrast, in 3D magnetic domain walls are  closed
surfaces on the dual lattice.  Specifically, if we write the expansion
%
\[
\Xi_N(u)= \sum_{n=0}^{Nd} b_n u^{2n}
~,
\]
%
then $b_n$ is the number of domain wall configurations of length $n$
(2D) or area $n$ (3D).  Taking into account the ground state energy
$-NdJ$, and a factor $2$ arising from the twofold degeneracy associated
with each domain wall configuration, the partition function is then
given in terms of  $\Xi_N$, as is well known~\cite{gattringer}:
\be
\label{eq-ghghghg}
Z_N= 2  u^{-Nd} ~ \Xi_N(u)~.
\ee

Per-site partition function and the other generating functions in
the thermodynamic limit $N\to\infty$ are defined as usual according to
\mbox{$
Z=
\lim_{N\to \infty} Z_N^{1/N}$}, \mbox{$
\Lambda=
\lim_{N\to \infty} \Lambda_N^{1/N}$} 
and 
\mbox{$
\Xi=
\lim_{N\to \infty} \Xi_N^{1/N}$.}


Crucially, 
given $\Xi_N$ one can obtain $\Lambda_N$ (and v.v.):
\begin{align}
  \Lambda_N(t)&= 2^{1-N} (1-t^2)^{Nd/2} u^{-Nd} ~\Xi_N(u)
  \nonumber
\\
&=  2^{1-N} (1+t)^{Nd} ~\Xi_N\left(\sqrt{1-t \over 1+t}\right) ~.
\label{eq-3489gt9rghwi4hgih}
\end{align}
In 2D,
both $\Xi$ and $\Lambda$ are explicitly known because
of Onsager's solution~\cite{onsager}.
But in 3D neither $\Xi$ nor $\Lambda$ is
explicitly known. 
However, there are partial results in 3D. The fermionic path integral
for $\Xi$ has been known~\cite{samuel3} since 1980.
In contrast, until now no
analogous expression for $\Lambda$ has been known. Our
main contribution here is to obtain the fermionic path integral for
$\Lambda$ for the simple cubic lattice, by using
(\ref {eq-3489gt9rghwi4hgih}).



\section{The Grassmann action for the Ising model}
\label{sec-action}

\subsection
{Grassmann variables and Berezin integration}
\label{sec-g}

The fermionic aspect of the 2D Ising model was already implicitly
apparent in the original works of Onsager and of Kaufmann, as seen
from their use of quaternion algebra~\cite{onsager} and generators of the
Pauli spin matrices~\cite{kaufman}.  A few decades later, in 1964,
Schultz, Mattis and Lieb formally showed that the 2D Ising model is
equivalent to a free-fermion model~\cite{lieb}, by employing 
fermionic creation and annihilation operators satisfying canonical
anticommutation relations. It was only in 1980 that the much more 
powerful fermionic path integral formulation of the 2D Ising model was
given~\cite{samuel1}  in terms of Grassmann variables, i.e.  fully
anticommuting quantities.

Let $\eta_i~(i=1,2,\ldots,N)$ be a set of Grassmann numbers that
satisfy
\be
\eta_i \eta_j +  \eta_j \eta_i =0 ~.  %
\ee
In particular, such quantities are nilpotent, $\eta_i^2=0$.  A general
power series in these $N$ quantities, with real or complex coefficients,
can thus only have $2^N$ terms at most, so that the Grassmann algebra
thus generated has dimension $2^N$.
Integrals of Grassmann variables are known as Berezin integrals, in honor of
Felix Berezin~\cite{berezin}, who showed how to modify Feynman's (bosonic) path
integrals to be applicable to fermions.
Berezin integration is a
translationally invariant linear operation defined
(in the standard convention used in physics)
according to
\begin{align}
  \displaystyle  \int \eta_i d\eta_i&~=~1 ~, \label{eq-berezin}
  \\ 
  \displaystyle \int d\eta_i \quad &~=~0 ~. %
  \end{align}
Multiple integrals can be defined as iterated integrals. 
Let $d\eta$ be shorthand for
$d\eta_1d\eta_2\ldots d\eta_N$  and 
$\eta$ for the entire  set $\{\eta_1,\eta_2,\ldots,\eta_N\}$. Consider
a general function 
\begin{align}
f(\eta)&= a_0
+ \sum_i^N a_i \eta _i
+ \sum_{i<j}^N a_{ij} \eta_i \eta_j
\nonumber 
\\
&  \quad \quad 
+ \cdots + 
a_{123\ldots N} \eta_1 \eta_2 \cdots \eta_N~.
\end{align}
Then from the definition of Berezin integration we find 
\be
\int f(\eta) d\eta = a_{123\ldots N}~. %
\ee
Hence,  Berezin integrals can be used retain only 
those terms that ``saturate'' the integral.  Using the Lagrangian path
integral formulation of Hamiltonian systems, a Berezin integral $\int
e^S d\eta$ of an exponentiated Grassmann action $S$ can be used to
select 
states with specific properties, rendering it an extremely
powerful tool in exact enumeration problems.


We will also require the use of a key property of how Berezin
integrals transform under changes of variables.  For usual
Riemann integrals $\int f(x) dx$ over, say, the reals $\Bbb R$, a
change of variables $x=ay$ with $a\in\Bbb R$ leads to $dx=ady$ for the
differentials.  However, 
for   Grassmann variables $x$ and $y$, if
$x=ay$ with $a\in\Bbb R$, then $dx= dy/a$ because of
(\ref{eq-berezin}).  In other words, the scaling  is in
the opposite sense.  Let us apply such scale transformations to
the actions, considered as 
functionals of the $\eta$.
Let  $q$ Grassmann
variables reside at each of $N$  lattice sites.
Consider the result of the dilation $\eta_i \mapsto \lambda \eta_i$, for
all $qN$ Grassmann variables. Let $\eta'=\lambda \eta$ be the rescaled
variables. Then applying the rule for  changing Grassmann variables, we obtain 
\mbox{$\int d\eta ~e^{S(\eta)}
  =
  \int d\eta' ~e^{S[\eta']}
  = \lambda^{-qN}
  \int d\eta ~e^{S[\lambda \eta]}$}.
Taking the $qN$-th root of $\lambda$  we thus get
\be
\label{eq-rescaling}
\lambda \int d\eta~ e^{S(\eta)}=    \int d\eta ~e^{S[\lambda^{1/qN} \eta]}~.
\ee
This renormalization of the Grassmann variables will play a central role
in the derivation below  of the exact  Grassmann action for 3D multipolygons.


\subsection
{2D multipolygons}
\label{sec-2d}

The original work of Samuel~\cite{samuel1}
included a remarkable
``one line'' solution of the 2D Ising model, leading immediately to
Onsager's solution~\cite{onsager} in less than a ``a page of
algebra''~\cite{samuel1}.  The generating function for 2D
multipolygons, whose coefficients are the celebrated series found by
Cyril Domb, also easily follows (see also  ref.~\cite{gmv}).  In notation similar to that of 
ref. \cite{gattringer}, the action for the isotropic model can be
written as
\begin{align}
  S_{2D}(\eta) &= u^2  S_L(\eta) + S_C(\eta) + S_M(\eta) \label{S-2D-erergerh}
  ~,  \\
  S_L(\eta)& = 
  \sum_{x\in \mathcal L}^N \bigg [\eta_{+1}(x) \eta_{-1}(x + \hat 1)
\nonumber
    \\
    & \quad  \quad  \quad \quad 
    + \eta_{+2}(x) \eta_{-2}(x + \hat 2) \bigg]
~,  \\
  S_M(\eta) &= \sum_{x\in \mathcal L}^N (\eta_{-1}(x) \eta_{+1}(x)  + \eta_{-2}(x) \eta_{+2}(x) )
~,  \\
  S_C(\eta) &= \sum_{x\in \mathcal L}^N \bigg[ \nonumber
  \eta_{+1}(x) \eta_{-2}(x)  + \eta_{+2}(x) \eta_{-1}(x) \\
& \quad  \quad  \quad \quad 
+  \eta_{+2}(x) \eta_{+1}(x)  + \eta_{-2}(x) \eta_{-1}(x) \bigg]~.
\end{align}
The sums are over all sites of the dual lattice $\mathcal L$ of the
magnetic domain walls.  The subscripts $M$, $L$, and $C$ denote
monomers, lines and corners~\cite{gattringer}.  Each line term is
associated with a single edge with weight $u^2$ of a single segment of
domain wall.  It can be checked by manual Berezin integration, for any
given configuration of lines, monomers and corners, that sites
attached to 1 or 3 lines give zero contribution, making the full
Berezin integral vanish.  Sites with 2 and 4 attached line terms
contribute with a factor $u^2$ for each line. Sites with no lines
contribute $-1$ but such sites can only ever appear in pairs, so 
for even $N$ the factor can be neglected.

The Berezin integral of the (exponentiated) action
(\ref{S-2D-erergerh}) thus enumerates multipolygons on the square
lattice, thereby counting all possible magnetic domain wall
configurations of the 2D Ising model.  Hence, 
\be
\Xi(u) = (-1)^N \int \prod^N_{x\in \mathcal L} d\eta_{-1}d\eta_{+1}d\eta_{-2}d\eta_{+2} \exp[S_{2D}]~.
\ee
We will write such expressions more compactly as
\be
\Xi(u) = (-1)^N \int d\eta \exp[S_{2D}]~.
\ee
%
The partition function is
then given by (\ref{eq-ghghghg}).  In the literature the factor
$(-1)^N$ is usually omitted because $N$ can be taken as even and,
moreover, it can be neglected in general when taking the thermodynamic
limit since $\lim_{N\to \infty}(-1)^{1/N} = 1$, but we  retain the sign for
completeness.
Fourier transformation and the application of the well known
determinant formula for Gaussian integrals immediately
leads~\cite{samuel1} to Onsager's solution (not shown here, see
Eq.~(3.12) in ref.~\cite{samuel1}).


\subsection
{The action for the 3D Ising model}
\label{sec-3d}

\begin{figure}[t]
~
  \includegraphics[width=0.95\linewidth]{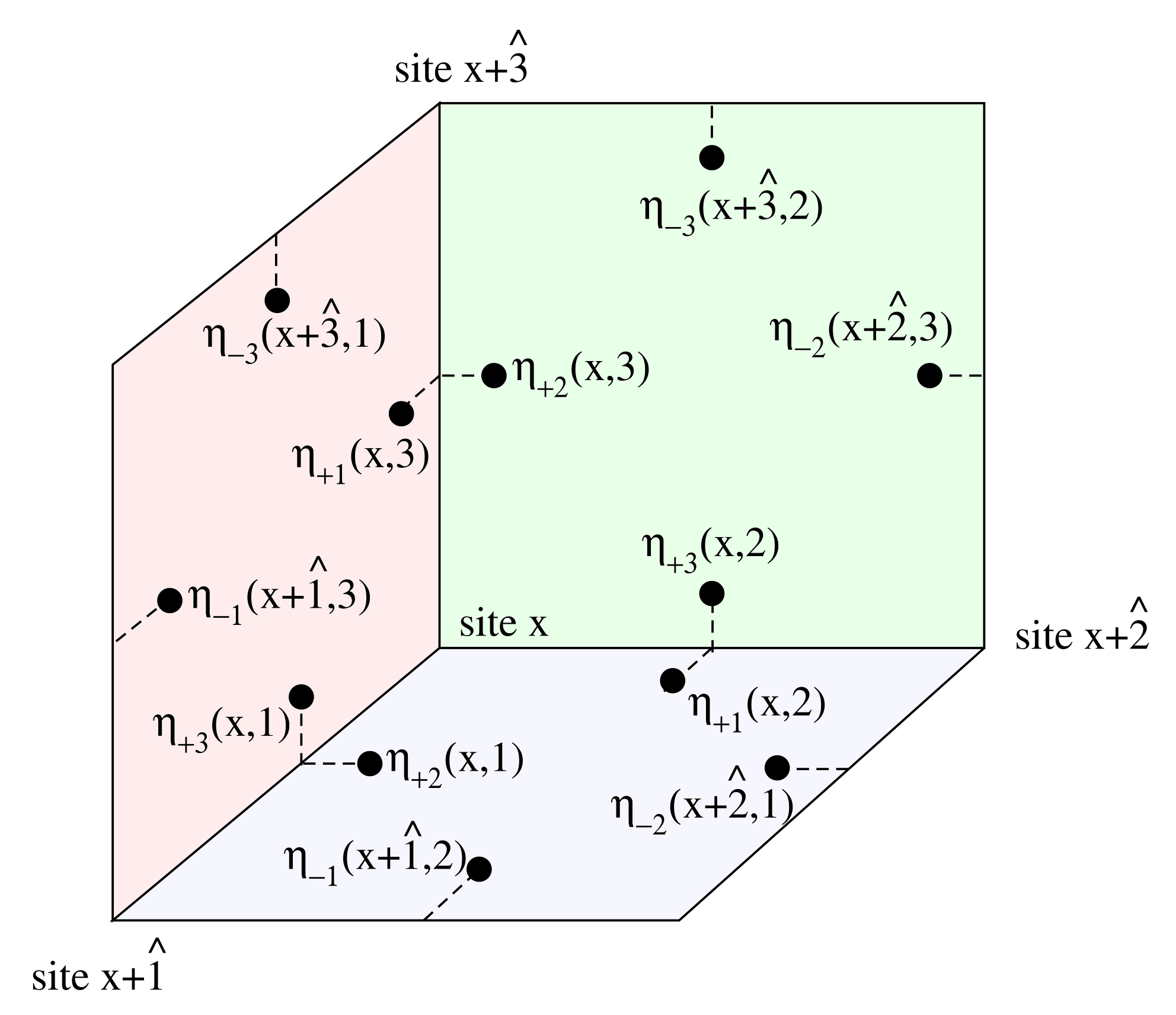}
 
\caption{Arrangement of the Grassmann variables on the (dual) 
  simple cubic lattice. The variables $\eta_{\pm \nu}(x,\mu)$ are
  indexed by the position $x$, by the type of edge $\mu$, and by
  positive or negative directions $\pm \nu$. The 3 primitive
  translation vectors are denoted $\widehat 1$, $\widehat 2$ and
  $\widehat 3$.  For a different perspective of the same
  arrangement, see Figure~6 of ref. \protect\cite{gattringer}.}
\label{fig}
  
\end{figure}

In 1980 Samuel also wrote down the
analogous Grassmann action for the 3D Ising model~\cite{samuel3}. However, this
action enumerates not multipolygons but rather closed surfaces.
Let $\eta$ denote the set of $12N$ Grassmann variables $\eta_{\pm\nu}
(x,\mu)$ following the notation of ref.~\cite{gattringer}, where $x$
is the site index as before and $\mu$ is the edge index, with $\pm
\nu$ describing the side of the edge  where the variable resides.  See
Figure \ref{fig} for the arrangement of the Grassmann variables.  Then
the action for the isotropic Ising model can be written as
\be
    S_{3D}(\eta,u)  = u^2~S_4(\eta) +S_2  (\eta)
    \label{eq-S}
    ~,
    \ee
where
\renewcommand\arraystretch{1.6}        
\begin{align}
 ~~   S_4 (\eta) &= \sum_x^N  
     \bigg\{ \nonumber             
\\ & 
    \begin{array}[t]{l}
      ~ ~  ~ 
\eta_{+2}(x,3) \eta_{-2}(x+\hat 2,3) \eta_{+3}(x,2) \eta_{-3}(x+\hat 3,2)
\\
+~
\eta_{+1}(x,3) \eta_{-1}(x+\hat 1,3) \eta_{+3}(x,1) \eta_{-3}(x+\hat 3,1)
\\
+~
      \eta_{+1}(x,2)
        \eta_{-1}(x+\hat 1,2) \eta_{+2}(x,1) \eta_{-2}(x+\hat 2,1) 
\bigg\} ~ ,  
 \end{array}
      \vspace{-2cm}~
    \label{eq-Sp}
\end{align}
    \begin{align}
         \nonumber
       S_2 (\eta) &= \sum_x^N 
      \sum_{\mu=1}^3 \sum_{\substack{~\nu,\rho\neq \mu \\ \nu<\rho}}^3
      \bigg\{  
      \begin{array}[t]{l} 
        \displaystyle
~~ ~ \eta_{+\nu}(x,\mu)~\eta_{-\rho}(x,\mu) \\
+~
\eta_{+\rho}(x,\mu)~\eta_{+\nu}(x,\mu) \\ 
+~
\eta_{+\rho}(x,\mu)~\eta_{-\nu}(x,\mu) \\
+~
\eta_{-\rho}(x,\mu)~ \eta_{-\nu}(x,\mu) \\
+~
\eta_{-\nu}(x,\mu)~\eta_{+\nu}(x,\mu) \\
+~
\eta_{-\rho}(x,\mu)~\eta_{+\rho}(x,\mu)
~\bigg\}~.~
      \end{array}
      \\  \label{eq-Sh}  
  \end{align}
   \renewcommand\arraystretch{1}        
 The quartic ``plaquette'' terms $u^2S_4$ and the quadratic ``hinge''
 terms $S_2$ above correspond to the terms $S_P(\eta)$ and
 $S_E(\eta)+S_M(\eta)$, respectively, in ref.~\cite{gattringer}.

As
with the 2D action, it is easy to see that this action
enumerates closed surfaces, as follows. Edges with 1 or 3 attached
plaquettes give contribution zero and render the full Berezin integral
zero.  Edges with 2 or 4 attached plaquettes contribute with a factor
$u^2$ for each plaquette. Edges with no plaquette give a factor of
$-1$.
Let
$d\eta$ be shorthand according to 
%
\[
d\eta=
\prod_x^N 
      \prod_{\mu=1}^3 \prod_{\nu \neq \mu }^3
      d\eta_{-\nu} (x,\mu) d\eta_{+\nu} (x,\mu)~.
      \]
%
Then the partition function is given by (\ref{eq-ghghghg}) with 
\be
\label{eq=-erbnjjgj}
\Xi_N (u)  = (-1)^{3N} \int d\eta \exp[S_{3D}]~.
\ee

Since this action is not quadratic, it does not correspond to a model
of free fermions, but rather to a model of interacting
fermions. Moreover, pfaffian and determinant formulas cannot be used
in the usual manner because the integrals are not Gaussian.
Nevertheless, over the decades significant progress been made even
without being able to obtain explicit expressions
and the Grassmannization research program has,
overall, been tremendously successful
~\cite{sI-2022,sI-01,sI-02,sI-03,sI-2020,sIII-01,sIII-03,sIII-02,grassmanization,sI-04,plechko,y1991,polyakov-1987}.

\section{The quartic action for enumerating 3D multipolygons}
\label{sec-results}

We now present our main results. 
In what follows, we will use the action (\ref{eq-S}) as a the starting point to
arrive at the analogous action for counting multipolygons on the
simple cubic lattice.
Substituting (\ref{eq=-erbnjjgj}) into (\ref{eq-3489gt9rghwi4hgih}) we get
\be
\label{eq-g449494}
\Lambda_N(t)=  2^{1-N} (1+t)^{3N} ~  (-1)^{3N} \int d\eta ~e^{S_{3D}(\eta,u)}~.
\ee
Applying the rescaling
(\ref{eq-rescaling}) to (\ref{eq-g449494}) we arrive at our first result:
\[
\Lambda_N(t)=  (-1)^{3N} \int d\eta ~e^{S_{3D}[ 2^{1/(12N)-1/12 } (1+t)^{1/4}   \eta,u]]}~.
 \]
  Let $S_{3DM}$ denote the 
Grassmann action for
exact enumeration of multipolygons on the simple cubic lattice. Then 
the above result can be written 
\begin{align}
\Lambda_N(t) &= (-1)^{3N} \int d\eta \exp[{S_{3DM}}] 
\\
  S_{3DM}(\eta,t) &= 
{S_{3D}( 2^{1/(12N)-1/12 } (1+t)^{1/4} \eta,u)}~.
\label{eq-343ygh}
\end{align}

Of particular interest in statistical mechanics are the thermodynamic
limits of various quantities, such as partition functions and
generating functions. Using (\ref{eq-S}) and  taking the  limit $N\to
\infty$, the (\ref{eq-343ygh})  simplifies to 
  \begin{align}
    S_{3DM}(\eta,t)
    &=
\frac{ (1+t)^{1/2}}{\sqrt[6]2}  S_2(\eta)  
+
\frac{(1-t)}{\sqrt[3]2} S_4(\eta) ~,
\label{eq-23545}
  \end{align}
  with $S_2$ and $S_4$ given by (\ref{eq-Sh} ) and (\ref{eq-Sp})
  respectively.

Finally, the claim that $S_{3DM}$ in (\ref{eq-23545}) is quartic
follows immediately from observing that $S_4$ by definition
(\ref{eq-Sp}) is quartic in the Grassmann variables.  Whereas
quadratic actions correspond to free fermion models, quartic actions
are associated with (typically unsolved) models of interacting
fermions.
The known singularity at $t=t_c$ of $\Lambda(t)$ for the simple cubic
lattice bears an important relation to string and gauge field theories
\cite{polyakov-1987}. {Indeed, it is possible to represent the
continuum limit of the 3D Ising model in terms of a fermion string
theory~\cite{polyakov-1979,dotsenko-1987,dotsenko-1988}.}

Note that 
$S_{3DM}$  is not polynomial
in the edge weight $t$, because Eq.~(\ref{eq-23545}) contains a square
root term that leads to a non-terminating binomial power series in
$t$. Hence, on the cubic lattice there are no well defined polynomial
``edge terms'' in the action, in contrast to the action $S_{2D}$ for 
the square lattice, which has the edge terms $S_L$ in
(\ref{S-2D-erergerh}).
Planar and nonplanar
polygons are very different indeed.

{ Nevertheless, note that upon expansion of the
  corresponding exponential and subsequent saturation of the Berezin
  integral, all surviving terms have an even number of quadratic terms
  contributing, such that the square root completely vanishes and the
  dependence on $t$ is again polynomial. Indeed, every plaquette
  contributes $(1-t)/\sqrt[3] 2$ and every ``missing plaquette''
  contributes precisely $(1+t)/\sqrt[3] 2$ in the thermodyamic limit.}

\section{Discussion and Conclusion}
\label{sec-end}

In summary, we have solved the 42-year-old problem of finding the
Grassmann action for exact enumeration of polygons on the simple cubic
lattice.  The Grassmann action for
enumerating multipolygons on the cubic lattice is quartic, not
quadratic, and has a remarkable non-polynomial dependence on the edge
weight $t$.
%
The significance of these results is that, on the simple cubic
lattice, enumerating multipolygons is of the same order of difficulty
as enumerating closed surfaces --- not easier.

{
  
  Nevertheless, it should be emphasized that there is
  no reason at all to expect this action to be unique. In the 2D case
  it is possible to give a constructive proof of this non-uniqueness,
  using the pfaffian or determinant formulas for the Gaussian
  integrals. In the 3D case the question is not so clear. Absent a
  mathematical proof, we cannot in principle completely rule out the
  existence of a different --- possibly even quadratic --- action that
  performs the same enumeration, however unlikely this may seem. These
  and similar issues merit further investigation.

Finally, we note that the results presented here suggest that
Grassmann actions can be found for polygon enumeration on diverse 
other regular lattices.}  We have preliminary results generalizing the above
results to other nonplanar lattices, which we hope to publish when
time permits.

~

\medskip

\section*{Acknowledgements}

We thank C.~G.~Bezerra, H. J. Jennings, A. M. Mariz
and J. H. H. Perk
for
comments and 
helpful feedback.
This work was supported 
by CNPq (grant numbers  \mbox{302051/2018-0} and \mbox{302414/2022-3}).

\end{document}